\documentstyle[epsfig]{aa}

\begin{document}

\def\etal{et al. }
\def\araa{Ann.\ Rev.\ Astron.\ Ap.}
\def\aplet{Ap.\ Letters}
\def\aj{Astron.\ J.}
\def\apj{ApJ}
\def\apjl{ApJ\ Lett.}
\def\apjs{ApJ\ Suppl.}
\def\aas{A\&A Supp.}
\def\aa{A\&A}
\def\aal{A\&A Lett.}
\def\mnras{MNRAS}
\def\mnrasl{MNRAS Lett.}
\def\nature{Nature}
\def\apss{Ap\&SS}
\def\pasa{{\it Proc.\ Astr.\ Soc.\ Aust.}}
\def\pasp{{\it P.\ A.\ S.\ P.}}
\def\pasj{{\it PASJ}}
\def\pre{{\it Preprint}}
\def\aph{Astro-ph}
\def\sovlet{{\it Sov. Astron. Lett.}}
\def\adspr{{\it Adv. Space. Res.}}
\def\expas{{\it Experimental Astron.}}
\def\ssr{{\it Space Sci. Rev.}}
\def\ar{{\it Astronomy Reports}}
\def\inpress{in press}
\def\inprep{in prep.}
\def\submit{submitted}

\def\ap{$\approx$ }
\def\mjysr{MJy/sr }
\def\inu{{I_{\nu}}}
\def\inufit{I_{\nu fit}}
\def\fnu{{F_{\nu}}}
\def\bnu{{B_{\nu}}}
\def\msol{{M$_{\odot}$}}
\def\mic{$\mu$m}
\def\cm2{$cm^{-2}$}


\title{Scanning strategy for mapping the Cosmic Microwave Background anisotropies with Planck}
\author{X. Dupac \& J. Tauber}
\institute{European Space Agency - ESTEC, Astrophysics Division, Keplerlaan 1, 2201 AZ Noordwijk, The Netherlands}

\offprints{xdupac@rssd.esa.int}

\authorrunning{Dupac \& Tauber}
\titlerunning{Scanning strategy for Planck}

\date{Received {} /Accepted {}}

\abstract{
We present simulations of different scanning strategies for the Planck satellite.
We review the properties of slow- and fast-precession strategies in terms of uniformity of the integration time on the sky, the presence of low-redundancy areas, the presence of deep fields, the presence of sharp gradients in the integration time, and the redundancy of the scanning directions.
We also compare the results obtained when co-adding all detectors of a given frequency channel.
The slow-precession strategies allow a good uniformity of the coverage, while providing two deep fields.
On the other hand, they do not allow a wide spread of the scan-crossing directions, which is a feature of the fast-precession strategies.
However, the latter suffer from many sharp gradients and low-coverage areas on the sky.
On the basis of these results, the strategy for Planck can be selected to be a slow (e.g. 4 month-period) sinusoidal or cycloidal scanning.
}

\maketitle


\section{Introduction\label{intro}}

In the last decade there has been a huge increase in the number of experiments dedicated to measurements of
the fluctuations of the Cosmic Microwave Background (CMB), e.g. Smoot et al. 1992, de Bernardis et al. 2000, Hanany et al. 2000, Kovac et al. 2002, Beno\^\i t et al. 2003, Pearson et al. 2003, Bennett et al. 2003b.
These experiments seek to measure a
signal whose intensity is very low with respect to the background signal ($\sim 10^{-5}$), over a wide range of angular scales, with the most
ambitious experiments aiming at covering the whole sky with an angular resolution of the order of 10$'$ or better. Such
experiments are plagued by a host of systematic effects which arise in the instrument itself (e.g. 1/f detector noise, thermal effects), or in celestial signals which are picked up by the instrument (e.g. straylight due to
Galactic emission entering through far sidelobes). 
A key factor in the success of these experiments is the pattern with which the detector lines-of-sight sweep
over the sky, often referred to as the scanning strategy.

The scanning strategy is usually defined on the basis of the instrumental capabilities, e.g. it is designed
to allow the suppression of inherent features in the detector behavior by the use of redundancies (or repeated
visits to the same part of the sky). This is akin in some ways to the use of hardware chopping, and is therefore largely specific to each experiment. At the same time the scanning strategy introduces symmetries in
the data acquisition stream, which has implications for the data reduction process which converts the time sequence of
acquired data into a map.
Since the data reduction is very computer-intensive (\cite{bond99}), the scanning strategy
design also attempts to enhance (respectively reduce) those symmetries which simplify (respectively complicate) the conversion.

In this paper we concentrate on the scanning strategy aspects of the Planck satellite, a third-generation CMB experiment; however there are generic features in our analysis which are applicable to other experiments with similar objectives.

The~Planck satellite (\cite{tauber01}, http://rssd.esa.int/Planck) is a European Space Agency project designed to map the whole sky at wavelengths ranging from 350 $\mu$m to 1 cm with unprecedented sensitivity (${\Delta T \over T} \approx 2 \times 10^{-6}$) and angular resolution (5$'$ - 30$'$).
Its main cosmological objective is to derive a very accurate angular power spectrum of the CMB fluctuations, both for the temperature and the E-mode polarization. It is also considered feasible that it
will detect and analyse B-mode CMB polarization signals.

The Planck payload consists of a large offset telescope which collects radiation from the sky and
delivers it to a focal plane populated with 70 detectors.
The detectors are of two kinds: tuned receivers based on low-noise amplifiers operating in the 20-80 GHz band
(Mennella et al. 2003), and wideband bolometers operating in the 90-1000 GHz range (\cite{lamarre03}). In both cases the detectors are fed by corrugated feedhorns which collect radiation from the telescope and define the angular responsivity of the detectors.
Both kinds of detector are very sensitive and stable, allowing them to operate in a ``total-power" mode. This was not the case for the earlier CMB experiments which today define the state of the art of CMB anisotropy research (COBE/DMR and WMAP, http://lambda.gsfc.nasa.gov); instead they were designed to be differential to overcome inherent instabilities in the detectors (i.e. the measured signals consist of differences of observations of different patches of the sky). The principle of Planck has however been demonstrated on smaller-scale balloon-based experiments, e.g.
BOOMERANG (\cite{debernardis00}) and Archeops (\cite{benoit02}).

In order to achieve its scientific objectives, the scanning strategy of Planck must ensure that: 
\begin{itemize}
\item the whole sky is covered by all detectors. An ideal survey would provide homogeneous sensitivity across the whole sky (indeed one of the most difficult issues in the data reduction process of large CMB surveys
is to understand the behavior of noise across the map), but this is in practice
very difficult to achieve by any continuous-sweep experiment. Given that there will be both under- and over-sampled regions of the sky, it is generically advantageous to confine them to compact, connected areas, which will offer specific scientific research opportunities (e.g. high-sensitivity and/or high-multipole CMB analysis, but also studies of the Sunyaev-Zeldovich effect).
\item there is enough redundancy at all time and angular scales to allow the detection and removal of spurious systematic effects down to the intrinsic sensitivity level of the detectors. The detection and removal process is largely ensured by processing algorithms often dubbed ``de-stripers" (e.g. \cite{delabrouille98}),
which are tailored to the specific experimental characteristics. The efficiency of the de-striper will also
depend on the type of spurious noises encountered; in particular map-making algorithms (e.g. Dor\'e et al. 2001, Natoli et al. 2001, Dupac \& Giard 2002) are very sensitive to non-stationary noises, and require crossing scans to be present in a large part of the sky.
\end{itemize}

The Planck satellite will be placed in a Lissajous orbit around the Sun-Earth L2 Lagrange point (see Fig.
\ref{orbitscan}), which guarantees low levels of thermal disturbances and straylight from the Sun, the Earth and the Moon.
The focal plane points at an angle of about 85$^o$ from the satellite's spin axis, so that as it rotates, the focal plane footprint (see Fig. \ref{focal}) traces a large circle on the sky.

\begin{figure}
\caption[]{Planck will be placed on a Lissajous orbit around the 2nd Lagrangian point of the Earth-Moon-Sun
system. The satellite spins slowly (1 rpm) around an axis which is pointed within 10$^o$ of the Sun-Earth line.
As it spins, the detector lines-of-sight sweep large circles around the sky.}
\label{orbitscan}
\end{figure}

\begin{figure}
\caption[]{The Planck focal plane projected on the sky sparsely samples a field of view of $\sim$8$^o$. The beam shapes, defined by each corrugated feed horn on the focal plane, are shown at the -20 dB levels.
Most horns feed radiation to two independent detectors sensitive to orthogonal linear polarisations; the crosses indicate the polarisation directions. When combined, each pair makes it possible to measure Stokes I and Q or U (\cite{tauber04}). Two pairs, rotated with respect to each other by 45$^o$ and aligned along the sweep direction, make it possible to measure I, Q and U. 
}
\label{focal}
\end{figure}

The satellite spin axis approximately tracks the direction of the Sun by means of regular repointing manoeuvres, cumulating to an average of 1$^o$ per day. These repointing manoeuvres must be such that
the sampling density of the sky in the repointing direction obeys the Shannon criterion (i.e. the spacing is at least 2.5 times smaller than the FWHM of the main beam patterns).
The scanning strategy is in practice the sequence of repointing manoeuvres that the satellite will carry out over its lifetime. 
A number of technical factors constrain rather severely the satellite attitude and must be taken into account when designing an optimal scanning strategy. Among these are:
\begin{itemize}
\item pointing accuracies
\item manoeuvrability and on-board fuel consumption
\item solar irradiation and thermal transient as well as periodic effects
\item contact with ground antennas and RF interference
\end{itemize}

More details on the constraints on the Planck scanning strategy can be found in Tauber (2003).

To determine an optimal scanning strategy for an experiment such as Planck, detailed modelling has to
be carried out which simulates: 
\begin{itemize}
\item the sources of signal (sky, but also spurious)
\item the convolution of the instrument response with the signals as a function of time
\item the conversion of the simulated time stream to maps
\item the detection and removal of unwanted signals to recover the CMB anisotropy map
\item ideally, also the compression of the CMB map into an angular power spectrum
\end{itemize} 
for each possible strategy, followed by a comparison of the noise properties of the recovered maps.
Such modelling is at least as complex as the processing foreseen on the acquired data, and prohibitive using traditional schemes
at the current time. A much simpler analysis is possible based on ``hit maps" (i.e. maps of the redundancy
behavior expressed as the number of visits per map pixel, or equivalently as the integration time per pixel)
and a relative assessment of these maps in each of the major areas previously mentioned.
In this paper we present such an analysis of the main scanning strategy options for Planck.

\section{Simulations of scanning strategies for the Planck mission\label{simus}}


For each scanning strategy considered, we have simulated the spin axis coordinates of Planck throughout a year,
and based on this we have produced integration-time maps of the whole sky with pixels of size 27$'$.
Strictly speaking, the resolution of these maps is not adequate for sampling the detector beams.
However, this does not affect our analysis, since we do not consider strategies which violate the sampling requirements
(see e.g. Sect. \ref{pre}), and any spatial structure
in the integration time at the scale of the beam size or lower is in reality smoothed out by convolution with the beam. 
For all maps we use the HEALPix Ring pixelization scheme of the sphere (\cite{gorski99}).
All simulations have been performed using home-made Interactive Data Language (IDL) procedures.

\subsection{The nominal strategy\label{nono}}

The simplest possible scanning strategy for Planck is one whereby it maintains the spin axis pointed as close
as possible to the Sun, while keeping strictly to the sampling requirement along the Ecliptic plane. 
This results in a displacement step of the spin axis along the Ecliptic plane of 2.5$'$ every hour. 
With this strategy, each detector observes the same (nearly 85$^o$-wide) circle $\sim$55 times
\footnote{the spin rate is 1 rpm, and the time required for each stepping manoeuvre is limited to 5 minutes.}
before moving to the next one (hereafter we shall refer to each spin revolution as a ``circle" and to each set of identical circles as a ``ring'').

This strategy (referred to as the nominal strategy) has the advantage that the inclination angle of the satellite to the Sun is kept nearly constant and orthogonal, which minimizes Sun-induced thermal variations. It also maximizes the mapping efficiency (i.e. minimizes the time lost in stepping manoeuvres).
A clear disadvantage is that most detectors will not survey the two five-degree radius caps around the Ecliptic poles; the fraction of the sky not seen varies from 0.008 \% (30 and 44 GHz) to 0.5 \% (143 GHz).
Because of its simplicity, the nominal strategy is the one used to assess the technical and operational capabilities of the satellite. It is however not likely that it will be used for Planck, due to the above
disadvantage.

Fig. \ref{nomiscan} shows the scan pattern of the main optical axis for a one-year survey, and Fig. \ref{nominal} shows the integration-time map corresponding to a one-year observation for one detector assumed to be on the main optical axis. The integration time is distributed in a simple manner with Ecliptic latitude,
peaking near the poles.

\begin{figure}
\caption[]{Scan pattern of the main optical axis for a one-year survey (nominal scanning strategy).
For clarity, only one ring in 300 is plotted.
The position of the spin axis is shown as a horizontal line.
}
\label{nomiscan}
\end{figure}

\begin{figure}
\caption[]{Integration-time map for one detector (located in the centre of the focal plane) and a one-year survey, for the nominal scanning strategy.
Mollweide projection of the whole sky in Galactic coordinates (the Galactic plane is horizontal, the Galactic north pole is at the top).
The grey areas at the Ecliptic poles indicate the non-observed regions.
}
\label{nominal}
\end{figure}

\subsection{Simple generic scanning strategies\label{pre}}

An excursion of the spin axis from the Ecliptic plane can be used to overcome the coverage holes left by
the nominal strategy around the Ecliptic poles. 
A second generic by-product of such excursions is to distribute the integration time to lower Galactic
latitudes than in the nominal case. This can be of benefit to a number of destriping algorithms (see
e.g. \cite{delabrouille98}) since it distributes the number of re-visits more widely over the sky.

However, the thermal and straylight design of the satellite requires that any excursions from the anti-solar pointing of the spin axis of the satellite must not have larger amplitudes than about 10$^o$. Furthermore, excursions 
entail the risk of generating thermal fluctuations which may result in systematic effects.

We have simulated two kinds of precession of the spin axis of the satellite:
\begin{enumerate}
\item in the first (``cycloidal") type, the spin axis follows a circular path around the anti-solar axis, which is equivalent to a cycloidal path as a function of Ecliptic longitude. Here the inclination angle of the satellite to the 
direction of the Sun is constant and therefore Sun-induced thermal variations are kept to a low level.
\item in the second (``sinusoidal") type, the spin axis path has two components: the first is along the Ecliptic plane, and the
second is perpendicular to the Ecliptic plane, with a sinusoidal amplitude in time.
Its path is therefore also sinusoidal along the Ecliptic equator.
\end{enumerate}

In both cases the period of the excursion is a free parameter; we have simulated both slow and fast periods for each kind.
The slow excursions have a period of 6 months, synchronised to the period of the orbit around L2,
which in the cycloidal case offers some advantages for the transmission of data to Earth.
It has however been shown by P. Leahy (priv. comm.) that it would be advantageous from the point of view of dipole-based calibration to reduce the period to 4 months.
We have performed simulations using a cycloidal precession with periods of 4 months and 2 months in order to characterize the intermediate strategies between the ``slow'' and ``fast'' ones.
The fast excursions are selected to have the shortest possible period (18 days) while providing a reasonable dwell time on each spin axis pointing (25 minutes); a shorter period would decrease the observing efficiency
to less than 80 \%.

Fig. \ref{nomiscan2} presents the scan patterns for a one-year observation in the case of the slow and fast cycloidal 7$^o$ strategies.
Fig. \ref{stratkinds} presents the corresponding observing-time maps for the cycloidal and sinusoidal, slow and fast, strategies, assuming a precession angle of 7$^o$.
Fig. \ref{angles} illustrates the effect of changing the amplitude of the excursion on the observing-time maps in the case of the slow cycloidal strategy.

\begin{figure}
\caption[]{Scan pattern of the main optical axis for a one-year survey with the slow (top) and fast (bottom) cycloidal 7$^o$ precessions.
For clarity, only one ring in 300 is plotted (one in 600 for the fast precession).
The positions of the spin axis are also shown.
}
\label{nomiscan2}
\end{figure}

\begin{figure}
\caption[]{Integration-time maps for one detector and a one-year survey, for, from top to bottom: the slow cycloidal, the slow sinusoidal, the fast cycloidal and the fast sinusoidal strategies.
The precession angle is 7$^o$.
Mollweide projections of the whole sky in Galactic coordinates (the Galactic plane is horizontal, the Galactic north pole is at the top).
}
\label{stratkinds}
\end{figure}

\begin{figure}
\caption[]{Integration-time maps for one detector and a one-year survey for the slow cycloidal strategy.
The precession angles are respectively 5$^o$ and 10$^o$ from top to bottom.
Mollweide projections of the whole sky in Galactic coordinates (the Galactic plane is horizontal, the Galactic north pole is at the top).
}
\label{angles}
\end{figure}

\subsection{Complex scanning patterns}

In principle any number of variations on the above simple types can be considered, as long as the basic attitude and sampling requirements are obeyed. Some of these variations are:
\begin{itemize}
\item large step-like variations in the spin-axis orientation
\item half-sine scans: sequential positive half-sine scans followed by negative ones
\item half-circular scans: idem but using circular paths
\item ``chaotic scans", which modifies the spin-axis orientation with random variations along an average path
\item nominal strategy plus specific scans to fill the holes at the Ecliptic poles
\item piece-wise combinations of different strategies
\end{itemize}

Our investigations into a few of these options indicate that they do not offer significant advantages over the
simpler strategies. Therefore we confine ourselves to an analysis of the latter.

\section{Results of the simulations}

We have analysed the various
scanning strategies in terms of properties of the distribution of the observation time on the sky.
Our main criteria are:
\begin{itemize}
\item the completeness and homogeneity of the coverage
\item the surface, depth and connectedness of deep fields
\item the~characteristics of sharp gradients in the integration-time maps
\item the redundancy in the scanning directions
\end{itemize}

In the following, we detail the characteristics of the different scanning strategies investigated towards these criteria, in terms of the response of a single detector located along the focal axis. 
In Table \ref{resu}, quantified characteristics are provided.
The (minor) effects of including all detectors in the analysis are addressed in Sect. \ref{coad}.

\subsection{Completeness and large-scale homogeneity}

The nominal strategy (Fig. \ref{nomiscan}) exhibits good uniformity and smooth coverage, except that the Ecliptic poles are not covered at all for some detectors, which is a shortcoming for this strategy. 
However, the non-covered area is relatively small (0.3 \% for a detector at the main optical axis), and it could be filled in by specific short-duration (a few days) observations.

As Fig. \ref{stratkinds} and \ref{angles} show, the slow precession strategies exhibit good uniformity of the integration time on the sky, except around the Ecliptic poles where we find deep-field candidates.
The fast-precession simulations show non-uniformities and sharp gradients over the whole map.
The fast-precession strategies have two significant weaknesses compared to the slow precession: the proportion of the sky which is not well covered is high (see Fig. \ref{histograms}), and there are no significant deep fields (see Sect. \ref{deep}).
The differences in terms of average integration time on the whole sky are due to the fact that more time is lost in manoeuvres in the case of the fast-precession strategies.

There are few differences between the cycloidal and sinusoidal strategies; one is that the sinusoidal strategies are perfectly symmetrical with respect to the Ecliptic equator, which is not the case for the cycloidal strategies.

The main difference that arises when changing the precession angle (see Fig. \ref{angles}) is that the size and depth of the long-integration patches change. Although not shown, this applies to sinusoidal strategies as well.

\begin{figure}
\caption[]{Distribution of the integration-time maps for one detector and a one-year survey for the slow cycloidal strategy (thin line) and the fast cycloidal strategy (thick line).
}
\label{histograms}
\end{figure}

\begin{table}[!ht]
\caption[]{\label{resu}
Characteristics of the simulated scanning strategies. N: nominal strategy; C5, C7, C10: cycloidal strategy with precession angle of 5$^o$, 7$^o$, 10$^o$; C7-4m, C7-2m: cycloidal 7$^o$ precessions with respectively a 4 months and 2 months period; S7: sinusoidal strategy with 7$^o$ precession angle; FC7, FS7: fast strategies with 7$^o$ precession angle, respectively cycloidal and sinusoidal; VFC7: very fast cycloidal precession (see Sect. \ref{continuous}).
The columns present the observing efficiency in percentage (Eff), i.e. the average integration time on the sky divided by the total time of the survey (1 year), the percentage of the sky which is not seen (NS), which has poor coverage (below half the average time: /2), good coverage (above twice the average: $\times$2), excellent coverage (above 4 times the average: $\times$4), the percentage of deep field surface on the sky (above 9 times the average time: DF), the percentage of surface suffering from sharp gradients ($>$ 2 in relative, G), the percentage of the sky covered by at least 3 scanning directions out of 4 spread in 45$^o$ (SD1), and the percentage of the sky covered by at least 5 scanning directions (SD2) out of 11 spread in 360$^o$ (see Sect. \ref{direct}).
The characteristics which may cause difficulties for a given strategy are marked in boldface.
}
\begin{center}
\begin{tabular}{llllllllll}
\hline
& Eff & NS & /2 & $\times$2 & $\times$4 & DF & G & SD1 & SD2\\
\hline
N & 92 & {\bf 0.3} & 4.6 & 5.4 & 1.4 & 0.2 & 0.7 & {\bf 1.2} & {\bf 0}\\
C5 & 92 & 0 & 6.2 & 5.3 & 1.0 & 0.3 & 0.8 & {\bf 2.5} & {\bf 0}\\
C7 & 91 & 0 &  8.4 & 4.6 & 1.2 & 0.4 & 0.9 & {\bf 3.5} & {\bf 0.3}\\
C7-4m & 91 & 0 & 6.8 & 4.9 & 2.4 & 0.5 & 1.3 & {\bf 5.2} & {\bf 0.3}\\
C7-2m & 91 & 0 & {\bf 32} & 9.7 & 4.4 & 0.3 & {\bf 2.5} & {\bf 10} & {\bf 2.5}\\
C10 & 91 & 0 & 12 & 4.3 & 2.2 & 0.4 & 1.0 & {\bf 5.1} & {\bf 0.5}\\
S7 & 92 & 0 & 2.4 & 4.6 & 1.2 & 0.5 & 0.8 & {\bf 3.7} & {\bf 0.2}\\
FC7 & 79 & 0 & {\bf 17} & 8.8 & 1.8 & {\bf 0} & {\bf 6.2} & 47 & 12\\
FS7 & 84 & 0 & {\bf 34} & 12 & 2.0 & {\bf 0} & {\bf 4.3} & 36 & 9.6\\
VFC7 & - & 0 & 0 & 6.1 & 1.9 & {\bf 0} & 0.6 & 60 & 19 \\
\hline
\end{tabular}
\end{center}
\end{table}

\subsection{The effect of co-adding several detectors\label{coad}}

Most detectors in the Planck focal plane are co-aligned along the scanning direction (Fig. \ref{focal}),
and for these the integration time maps scale up simply from that of a single detector.
For many frequency channels, however, the detectors are spread more widely across the scan direction and therefore some
smoothing effect can be expected when co-adding all detector maps at a given frequency.
In Table \ref{alldet} we illustrate this effect in the case of the slow cycloidal strategy.
The generally small differences seen do not change the conclusions drawn from the analysis of one detector.
Although not explicitly shown, for other types of strategy the situation is similar.

\begin{table}[!ht]
\caption[]{\label{alldet}
Coverage characteristics of the Planck channels when co-adding all detectors, for the slow cycloidal precession (C7).
The columns show the Planck frequency channel (GHz), the percentage of the sky which has poor coverage (below half the average time), good coverage (above twice the average), excellent coverage (above 4 times the average), and the percentage of deep field surface on the sky (above 9 times the average time).
}
\begin{center}
\begin{tabular}{lllll}
\hline
& $<$ avg/2& $>$ avg$\times$2& $>$ avg$\times$4& deep fields\\
30 & 12.3 & 4.2 & 1.5 & 0.43 \\
44 & 6.2 & 5.3 & 1.6 & 0.28 \\
70 & 9.3 &4.3 & 1.6 & 0.48\\
100 & 9.6 & 4.4 & 1.4 & 0.44\\
143 & 6.9 & 4.9 &1.1 & 0.51\\
217 & 8.0 & 4.5 & 1.4 & 0.47 \\
353 & 8.4 & 4.6 & 1.2 & 0.45\\
545 \& 857 & 8.2 & 4.8 & 1.1 &0.45\\
\hline
\end{tabular}
\end{center}
\end{table}

\subsection{Deep fields\label{deep}}

As discussed previously, areas with very high signal-to-noise are of special scientific interest.
It should however be noted that the Planck design implies that these fields will inevitably be located at high (positive and negative) Ecliptic latitudes, which are certainly not the cleanest parts of the sky in terms of foregrounds.

We define ``deep fields" as areas that have at least 9 times more integration time than the sky average (i.e. 3 times more sensitivity).
It is clear that almost any slow scanning strategy allowed for Planck will produce deep fields.
As can be seen in Table \ref{resu}, the two slow-precession strategies make it possible to obtain two relatively large deep fields (about 90 deg$^2$ each) around the Ecliptic poles.
It is also important that the areas with high signal-to-noise do not consist of many disconnected patches.
This is the case for the slow strategies (see Fig. \ref{zooms}). The effect of
changing the amplitude of the excursion (Fig. \ref{angles}) is to spread the highly-covered fields (avg$\times$4) over a larger area, but the size of the really deep fields is not much affected.

Additional issues (Leahy, priv. comm.) to be considered with respect to deep fields are:
\begin{itemize}
\item the extent to which deep fields overlap for different frequency channels.
This is adequate at least for the central (CMB) channels.
\item the connectivity within the deep fields. In this respect, 6-month period strategies are better than shorter-period strategies where the symmetries imply that the fields break up into several patches.
\end{itemize}

\begin{figure}
\caption[]{Close-up of the North Ecliptic polar regions for the slow (top) and fast (bottom) cycloidal strategies.
The South Ecliptic regions are similar to these.
}
\label{zooms}
\end{figure}

\subsection{Sharp gradients in the observing-time map}

The presence of sharp angular gradients in the noise characteristics creates severe difficulties for
the recovery of the wanted signals (map-making). Detailed simulations are required to determine
what levels are acceptable, furthermore currently the exact map-making algorithms that will be used
by Planck have not been selected. Therefore we define a conservative threshold of acceptability
as being that where the ratio between the integration-time gradient (per pixel) and the time quantity is below 2. 
Table \ref{resu} presents the fractional sky area which exhibits unacceptably sharp gradients, 
according to the above definition, and calculated using the Roberts edge-detection technique.

This analysis indicates that the fast scanning strategies exhibit a non-negligible amount of areas
with sharp gradients.
In contrast, the nominal and slow strategies do not exhibit a significant amount of such areas.

\subsection{Scanning directions per pixel\label{direct}}

The measurement of polarisation presents specific challenges for the scanning strategy. As outlined
previously, to measure Stokes I, Q and U, Planck makes use of four detectors which are aligned along
the scan direction. Together, the four detectors probe four directions of linear polarisation and constitute
a minimum necessary set. However, probing more directions would help considerably to pin down more
accurately the recovery errors.
To be effective, the directions probed by each of the detectors thanks to the scanning redundancy should be homogeneously spread between 0$^o$ and 45$^o$ (\cite{couchot99}), e.g. if only two measurements are taken, they should ideally be 22.5$^o$ apart. The focal plane design of Planck is such (see Fig. \ref{focal}) that it is enough to use one detector (out of each co-aligned group of four) to gauge the number of directions probed, and to use as a reference the
direction of the scan across the sky (rather than the actual polarization direction of the detector).
Therefore, we calculate the scan angle of each observation, i.e. the angle the scan direction makes with respect to isolatitude lines, and we count the number of visits per pixel per scan direction, collecting them into 4 bins spread over 45$^o$ (each scanning angle $\alpha$ is equivalent to $\alpha$ modulo 45$^o$).
Table \ref{resu} presents the percentage of the sky which is well covered in terms of scanning directions within 45$^o$ (3 or 4 directions).
These numbers make it possible to have an idea of how good the redundancy is for the determination of the Stokes parameters, with respect to the different scanning strategies.

A different issue is the detection and removal of systematic effects which depend on the polarization direction.
In this respect, it is important to have as many crossings as possible for each detector (e.g. \cite{revenu00}), at least in some parts of the sky where it would be possible to understand very well the behaviour of the detectors.
To be effective, the directions probed should be widely spread around 2$\pi$.
It should also be pointed out that some direction-dependent systematic effects may average out in areas with widely spread directions.
The last line of Table \ref{resu} presents the percentage of the sky which is well covered in terms of scanning directions around 2$\pi$, i.e. at least 5 bins out of 11 have been visited by a given detector.
This percentage of the sky is not distributed over the sky as homogeneously as the total integration time,
as is shown in Fig. \ref{scandir}.

\begin{figure}
\caption[]{The distribution of the number of scanning direction bins probed by one detector in each pixel, in the case of the slow (top) and fast (bottom) cycloidal strategies.
The sampled directions are collected in 11 bins evenly spread around 2$\pi$.
Mollweide projections of the whole sky in Galactic coordinates (the Galactic plane is horizontal, the Galactic north pole is at the top).
}
\label{scandir}
\end{figure}

Fig. \ref{nomiscan} and \ref{nomiscan2} also indicate the range of different crossing angles for each strategy.
It is clear that none of the nominal and slow-precession strategies fulfil the desirable requirements in terms of range of directions probed.
Indeed, a negligible part of the sky is covered with at least 5 scanning directions over 11, and the angles between directions are small in most areas.
The fast strategies are clearly much better in this respect.
The same trend is observed for 45$^o$-spread bins, which shows that fast-precession strategies are superior both for measuring the polarization and for removing the systematics.

\section{Comparison with continuous-precession strategies\label{continuous}}

The choice of an optimal scanning strategy for WMAP has followed a rather different rationale than for Planck.
Indeed, WMAP (\cite{bennett03a}) uses a very fast (1 hr$^{-1}$) and continuous precession of the spin axis, and therefore scans a large part of the sky in a short time. Planck uses a slow step-wise
motion which emphasizes revisits on short timescales and makes it possible to achieve very high sensitivity over a much
smaller area per day. These different strategies can be traced to the choice of a differential (WMAP) versus
``total-power" detector design, which was in turn driven by the original expectations of detector stability in
each case, and the design philosophy.
The WMAP approach requires a very complex optical design (dual telescope, matched detectors), whereas in the case of Planck the optics are simpler and make it possible to focus the design complexity in other areas (e.g. large focal plane arrays, maximising sensitivity; pointing accuracy and knowledge; but also cryogenics, which allows the use of extremely sensitive detectors - bolometers).
In addition, the single-telescope design of Planck allowed more flexibility in designing the baffle which shields the optics from the Galactic straylight and provides sharp edges in the detector responsivity at large angles, a feature which can be efficiently used to detect systematics (e.g. \cite{delabrouille98}).

A differential approach {\it \`a la} WMAP implies a time-stream with very reduced long-time-scale correlations (e.g. \cite{tegmark97}), and
in this situation the best map-making algorithms require to estimate the low-frequency noise components from
both time- and spatial-domain constraints (\cite{stompor03}). At the same time, a manageable amount
of data implies that it is possible to apply optimal methods to the
map-making problem. These characteristics imply that the best strategy for WMAP is indeed to uninterruptedly cover a large
amount of area in a short time, maximizing the number of scan crossings (spatial constraints) at all
time-scales.

In the case of Planck, the very large amount of data to be converted into maps would complicate the use of optimal map-making algorithms in the case of a continuous-precession strategy; rather, two-step methods have to be used
which require the pre-estimation of noise characteristics. Among the simplest of such methods are the
de-striping algorithms (e.g. \cite{delabrouille98}) which estimate the low-frequency noise components based on
the pre-processing of multiply-observed (high-sensitivity) scan paths.
Such algorithms have been shown to be very effective when applied to Planck.
The ``total-power'' detector design of Planck (which does produce time-streams with significant levels of low-frequency noise), and its map-making requirements, are therefore fully consistent with the current approach based on
repeated circles which almost achieve the final sensitivity in the observed pixels each hour.
Nonetheless, the level of scan cross-linking is clearly not as widely spread on the sky as could be wished, even when making use of the flexibility allowed by the spin-axis excursion strategies explored in Sect. \ref{simus}.
This is particularly true if unexpected systematic signals arise with time-scales of order one day
(the time scales where scan cross-linking is most deficient for Planck).

To overcome some of these difficulties, K. G\'orski, and C. Lawrence (2000, priv. comm.), advocated the implementation for Planck of a continuous precession of the spin axis with a period of around one day.
Wright (2004) also suggested that the Planck scanning strategy would be much better if its excursion period was of the order of 10 hours.
While continuous precession has proved impossible to implement due to the inertial behavior of the Planck spacecraft, it could in principle be approximated by the satellite in a step-wise fashion, though it
would most probably not be compatible with a two-step map-making approach and may also lead to sampling problems for
the highest-frequency detectors; in addition the observing efficiency would drop dramatically and the fuel
consumption rise to unacceptable levels. In spite of these difficulties which make it impossible to implement this approach, we have simulated the coverage that such a {\it hypothetical} scenario would lead to.
In this thought experiment, we
keep the nominal feature of having one 85$^o$-wide ring per minute, shifted successively by 2.5$'$ per minute
in a 7$^o$ radius circle around the spin axis, leading to a full excursion period of 17.6 hours.
This period would be the absolute minimal value possible given the sky-sampling requirements of the highest
frequency detectors.
Fig. \ref{contiscan} presents the scan pattern obtained from this ``very fast precession strategy", in the case of a cycloidal path.
Fig. \ref{conti} shows the integration-time map resulting from this strategy, and its characteristics are
presented in Table \ref{resu}.

\begin{figure}
\caption[]{Scan pattern of the main optical axis for a one-year survey (very fast cycloidal scanning strategy).
For clarity, only one ring in 20000 is plotted.
}
\label{contiscan}
\end{figure}

\begin{figure}
\caption[]{Integration-time map for one detector and a one-year survey for the very fast cycloidal-precession strategy.
The precession angle is 7 degrees.
Mollweide projection of the whole sky in Galactic coordinates (the Galactic plane is horizontal, the Galactic north pole is at the top).
}
\label{conti}
\end{figure}

This very fast precession would allow a very good uniformity of the integration time on the sky, as can be seen from Table \ref{resu}, in this sense it is not too different from the nominal strategy studied in
Sect. \ref{nono}.
Its main drawback is that no deep fields are present.

The number of crossings is not dissimilar to that of the other strategies, but they are much more
evenly spread across the sky, in particular to lower Ecliptic latitudes.
The crossings are also present at short time scales, which is clearly advantageous from the point of view of map-making.
This strategy is moreover much better than any of the others in terms of the coverage of scanning directions per pixel.

\section{Conclusion}

We have presented simulations of the observing-time distribution on the sky for a variety of scanning strategies for the Planck surveyor.
The slow-precession strategies are better in general than the fast-precession ones, because they allow a good uniformity of the coverage, while providing two deep fields.
On the other hand, they do not allow a wide spread of the scan-crossing directions.
This drawback will become a serious problem for the mission requirements only if the systematic effects are much worse than expected.
This characteristic can be improved by reducing the precession period to 2 months, which would also favor the dipole-based calibration.
However, the 2 month-period strategy suffers from bad coverage in a large part of the sky and many sharp gradients.
A 4-month period does not induce such drawbacks, but the gain in scanning directions with respect to 6 months is negligible.
It would however be better in terms of dipole-based calibration.

No very significant differences are found between cycloidal and sinusoidal simulations, or indeed with any other smooth precession for a given amplitude and period.
The sinusoidal strategies offer a slight advantage in terms of north/south uniformity, and the cycloidal one in terms of decreased response to thermal fluctuations.

A continuous precession of the spin axis would allow a good uniformity of the coverage and a good redundancy in the scanning directions.
However, this last option is not technically feasible given the Planck mission design.

In this context, a reasonable choice for a baseline strategy for Planck will be driven by tests of thermal response to be performed in the early part of the mission.
If the response is very low, the choice will be for a sinusoidal strategy, probably with a precession period of four months.
In the opposite case, a cycloidal strategy will be preferred.
This preliminary selection will be confirmed in the coming years by on-going simulations which will examine the impact of various scanning strategies on the noise covariance matrix of recovered CMB maps.

\section{Acknowledgements}

We would like to thank K.M. G\'orski and his collaborators for their so
useful HEALPix package.
We also thank very much J.-P. Bernard for having provided us with seed software, J.-L. Puget for very useful comments and many other members of the Planck collaboration for fruitful discussions.


\begin{thebibliography}{}

\bibitem[Bennett et al. 2003a]{bennett03a} Bennett, C.L., Bay, M., Halpern, M., et al.: 2003, a, ApJ, 583, 1

\bibitem[Bennett et al. 2003b]{bennett03b} Bennett, C.L., Halpern, M., Hinshaw, G., et al.: 2003, b, \apjs, 148, 1

\bibitem[Beno\^\i t et al. 2003]{benoit03} Beno\^\i t, A., Ade, P., Amblard, A., et al.: 2003, \aal, 399, L19

\bibitem[Beno\^\i t \etal 2002]{benoit02} Beno\^\i t, A., Ade, P., Amblard, A., et al.: 2002, Astrop. Phys., 17, 101

\bibitem[Bond et al. 1999]{bond99} Bond, J.R., Crittenden, R.G., Jaffe, A.H., Knox, L.: 1999, Comput. Sci. Eng., 1, 21

\bibitem[Couchot et al. 1999]{couchot99} Couchot, F., Delabrouille, J., Kaplan, J., Revenu, B.: 1999, A\&A Supp., 135, 579

\bibitem[de Bernardis \etal 2000]{debernardis00} {de Bernardis}, P., Ade, P.A.R., Bock, J.J., \etal:
  2000, \nature, 404, 955

\bibitem[Delabrouille 1998]{delabrouille98} Delabrouille, J.: 1998, A\&A Supp., 127, 555

\bibitem[Dor\'e \etal 2001]{dore01} Dor\'e, O., Teyssier, R., Bouchet, F.R.,
  Vibert, D., Prunet, S.: 2001, \aa, 374, 358

\bibitem[Dupac \& Giard 2002]{dupac02} Dupac, X., Giard, M.: 2002, MNRAS, 330, 497

\bibitem[G\'orski \etal 1999]{gorski99} G\'orski, K.M., Hivon, \'E., Wandelt, B.D.: 1999, proc. of the MPA - ESO cosmology conf., Garching, Germany, 2-7 Aug. 1998, A.J. Banday, R.K. Sheth, L.N. da Costa, eds, IPSKAMP NL, p. 37

\bibitem[Hanany \etal 2000]{hanany00} Hanany, S., Ade, P., Balbi, A., \etal: 2000, \apj, 545, L5

\bibitem[Kovac et al. 2002]{kovac02} Kovac, J.M., Leitch, E.M., Pryke, C., Carlstrom, J.E., Halverson, N.W., Holzapfel, W.L.: 2002, Nature, 420, 772

\bibitem[Lamarre et al. 2003]{lamarre03} Lamarre, J.-M., Puget, J.-L., Bouchet, F., et al.: 2003, New Ast. Rev., 47, 1017

\bibitem[Mennella et al. 2003]{mennella03} Mennella, A., 
Bersanelli, M., Seiffert, M., Kettle, D., Roddis, N., Wilkinson, A., \& 
Meinhold, P.\ 2003, \aa, 410, 1089 

\bibitem[Natoli et al. 2001]{natoli01} Natoli, P., de Gasperis, G., Gheller, 
C., \& Vittorio, N.\ 2001, \aa, 372, 346 

\bibitem[Pearson et al. 2003]{pearson03} Pearson, T.J., Mason, B.S., Readhead, A.C.S., et al.: 2003, ApJ, 591, 556

\bibitem[Revenu et al. 2000]{revenu00} Revenu, B., Kim, A., 
Ansari, R., Couchot, F., Delabrouille, J., \& Kaplan, J.\ 2000, A\&A Supp., 142, 
499 

\bibitem[Smoot \etal 1992]{smoot92} Smoot, G.F., Bennett, C. L., Kogut, A., \etal: 1992, \apjl, 396, L1

\bibitem[Stompor \& White 2003]{stompor03} Stompor, R., White, M.: 2004, A\&A, 419, 783

\bibitem[Tauber 2001]{tauber01} Tauber, J.A.: 2001, {\it The Extragalactic Infrared Background and its Cosmological Implications}, proc. of IAU Symp. 204, held 15-18 Aug. 2000, in Manchester, UK, ed. M. Harwit, p. 493

\bibitem[Tauber 2003]{tauber03} Tauber, J.A.: 2003, {\it Planck Scanning Strategy Reference Document}, ESA

\bibitem[Tauber 2004]{tauber04} Tauber, J.A.: 2004, The 
Magnetized Interstellar Medium, proc. of the conf. held in 
Antalya, Turkey, Sept. 8-12, 2003, eds: B.~Uyaniker, W.~Reich and 
R.~Wielebinski, Copernicus GmbH, Katlenburg-Lindau., p.~191

\bibitem[Tegmark 1997]{tegmark97} Tegmark, M.: 1997, Phys. Rev. D, 56, 4514

\bibitem[Wright 2004]{wright04} Wright, E.L.: 2004, {\it NATO ASI on "Frontiers of the Universe: Cosmology 2003"} in Carg\`ese, Sept 2003, astro-ph/0401001

\end{thebibliography}
\end{document}